\pdfoutput=1
\documentclass{PoS}
\usepackage{amsmath}
\usepackage{graphics}

\title{Quarks and quantum statistics}

\ShortTitle{Quarks and quantum statistics}
\author{\speaker{O.W. Greenberg}\footnote{University of Maryland Preprint PP-11-001}\\
Center for Fundamental Physics, Department of Physics\\
  University of Maryland, College Park, MD~~20742-4111, USA\\
and Helsinki Institute for Physics\\
P.O.Box 64 (Gustaf Hällströmin katu 2)\\
FIN-00014 University of Helsinki, Finland\\
E-mail: \email{owgreen@umd.edu}}

\abstract{I write about H\'ector, his contributions to the early work in the quark model, and a general
discussion of quantum statistics }

\FullConference{Quarks, Strings and the Cosmos - H\'ector Rubinstein Memorial Symposium\\
		August 09-11, 2010\\
		AlbaNova )Stockholm) Sweden}

\begin{document}

\def\beee{\begin{equation}}
\def\eeee{\end{equation}}
\def\dggg{^{\dagger}}
\def\bf{\mathbf{}}

\bibliographystyle{unsrt}

\section{Introduction}\label{introduction}

H\'ector and I first met at a summer workshop in Bebek, near Istanbul, Turkey.
The workshop photograph is at the end of this paper.
At first sight, H\'{e}ctor is nowhere to be seen.
Actually, he is there, but he is turned around talking or arguing with somebody
behind him. That is H\'{e}ctor, too full of energy to pose passively for a photo.

The tallest person above the left
of the last row is Shelly Glashow, to his left is Nick Burgoyne. One row down, from the
left, are Arthur Jaffe, Joe Dothan, Sidney Coleman, Gian-Fausto Dell'Antonio, Bruria
Kaufman, and Eduardo Caianiello, among others. On the left of the third row down is
H.R. Mani, and on the right of that row, there is H\'ector, turned facing backwards as just mentioned.
David Fairlie is near the left end  of the fourth row. Eugene Wigner is near the left and
Giulio Racah is near the right end of the next row. Louis Michel is a bit to the right
of the bottom row, Feza G\"ursey, the organizer of the workshop, is at the end, and I am
sitting next to him.

H\'{e}ctor and I stayed in touch after the Bebek workshop. We met often in Paris,
where H\'{e}ctor liked to go to Cafe Select Latin, which was frequented by people
from Latin America. We also met in Moscow, Helsinki, Stockholm, and in Maryland,
where for some years H\'{e}ctor made an annual visit, and gave the denizans of
College Park the news from Israel and Europe.

H\'ector was brusque, outspoken, stimulating, energetic, fun to be with, and
overall one-of-a-kind, honest to a fault, and full of vitality. We all miss him.

\section{Outline of my talk}\label{outline}

The first part of my talk will review the early days of the quark model, and
H\'ector's early contributions. Then I will shift to a general discussion of
quantum statistics with emphasis on two topics: the possible statistics that quantum
mechanics and quantum field theory allow, and the precision experimental tests
for the electron and the photon. I will conclude with some brief remarks about
H\'ector's last interests--cosmology and astroparticle physics.

\section{Quarks and aces}\label{quarks and aces}

As we all know, the quark model was introduced by George Zweig~\cite{z} and Murray Gell-Mann~\cite{mgm}
in 1964. Zweig focussed on the constituent aspect of what he called ``aces.''
Gell-Mann focussed on the algebraic properties of his ``quarks.''

Zweig was struck by the dominance of $\phi \rightarrow K \bar{K}$
over $\phi \rightarrow \rho \pi$. Experiment showed that
$\phi \rightarrow \rho \pi$ was suppressed by over two orders of magnitude.
This was an important motivation for Zweig. Zweig explained this by making
the strange and the non-strange mesons from different aces. In doing this
Zweig had to disregard Feynman's dictum that ``in the strong interactions everything that
can possibly happen does, and with the maximum strength allowed by unitarity.''
Gell-Mann's version of this rule was the ``totalitarian principle'' that
``everything that is not forbidden is compulsory.'' Zweig drew quark line diagrams,
``Zweig diagrams,'' to illustrate his rules. Allowed processes must have some
quark lines connecting the initial and final states.

To make baryons out of 3 aces, Zweig had to assign baryon number $1/3$ to his aces.
This led directly to fractional electric charges for the aces. Zweig found mass
relations, $m_{\omega}^2=m_{\rho}^2$ and
$m_{K^{\star}}^2=(m_{\phi}^2 + m_{\rho}^2)$ that agreed well with experiment.
Now we understand that Zweig's aces are the constituent quarks of the quark model.

Gell-Mann's path to quarks went through models from which he abstracted symmetries.
He went from global symmetry to $SU(3)$ (his ``eightfold way'').
He introduced ``leptons'' $L$ and $l$ as a tool to construct unitary spin.
In his 1964 paper he proposed the quark model with fractional electric charges.
In his quark model paper, Gell-Mann wrote~\cite{mgm} ``It is fun to speculate about the way
quarks would behave if they were physical particles of finite mass (instead of purely
mathematical entities as they would be in the limit of infinite mass). ... ''
He continued with ``A search ... would help to reassure us of the non-existence of real quarks.''
Gell-Mann used the method of \textit{abstraction} from a Lagrangian field theory. He used the
symmetries of the Lagrangian to infer symmetries of the theory. In another paper he wrote,
``We compare this process to a method sometimes employed in French cuisine: a piece
of pheasant meat is cooked between two slices of veal, which are then discarded.''

Feza
G\"ursey and Luigi Radicati~\cite{gur}, were the first to introduce spin in the quark model in an intrinsic way.
Choosing spin $1/2$ for the quarks is the simplest way to account for baryons of spin $1/2$.
G\"ursey and Radicati went much further. They unified the $u,~d,~s$, ~$SU(3)_{flavor}$
symmetry, with spin up and spin down,~ $SU(2)_{spin}$ symmetry, to $SU(6)_{flavor-spin}$
symmetry in which the quark is a $6$. With this larger symmetry they
recovered the Gell-Mann--Okubo mass relations for the baryon octet and decuplet and
found a new relation between the masses of the octet and decuplet. Mirza A. Baqi B\'eg,
Benjamin Lee and Abraham Pais~\cite{beg} calculated the magnetic moment ratio, $\mu_{p}/\mu_{n}$,
using this symmetry,
another success. G\"ursey and Radicati assigned the ground state baryons to the symmetric quark state.
This assignment was paradoxical, because it violated the Pauli exclusion principle.

G\"ursey and Radicati presented the $SU(6)_{flavor-spin}$ theory as a step towards
a relativistic theory. Several authors explored higher symmetries, such as
$U(6,6)$, $\tilde{U}(12)$, etc., in order to make
$SU(6)$ a fully relativistic theory. Some physicists hoped that a fully relativistic
version of $SU(6)$ would wrap up particle physics.
No-go theorems, culminating in Coleman and Mandula~\cite{col} and in Haag, Lopuszanski and Sohnius~\cite{haa}
proved that the only way to incorporate an internal symmetry with Poincar\'{e} symmetry
is supersymmetry. That was not to be the way.

What worked was to
take quarks as real, but change their statistics.
I realized that the Pauli exclusion principle is valid if quarks are parafermions of order 3.~\cite{owg}
I predicted the spectrum of excited states of baryons using parafermi quarks of order 3
and the fact that 3 quarks can be in a symmetric state
in the visible degrees of freedom, space, spin and flavor, with the antisymmetry needed for
the Pauli principle accounted for by the parafermi statistics,

Here the 3 of parafermi statistics is the same 3 as the 3 of color.
Parafermi statistics of order 3 is
equivalent to color $SU(3)$ as a classification symmetry.

\section{Mixed response of physics community}\label{mixed responce}

The response of the physics community to quarks or aces was mixed.
Zweig took them as real. Gell-Mann's attitude was ambivalent.
Gell-Mann's response to Zweig when Zweig returned from CERN was ``Oh, the concrete quark model.
That's for blockheads!''~\cite{z2} Many physicists were deeply skeptical of the quark
model, and even more of color.
Oppenheimer's response to me when I showed him my paper in 1964 was ``Greenberg, it's beautiful,
but I don't believe a word of it.''
Steven Weinberg wrote~\cite{wei}, ``At that time (1967) I
didn't have any faith in the existence of quarks.''

In contrast to Gell-Mann, Oppenheimer, Weinberg and many other physicists, H\'ector
wholeheartedly embraced the idea of quarks as real objects. He immediately started
developing models that required quarks to be physical particles, not mere mathematical constructs.
In the one year, 1966, H\'ector wrote or co-authored 6 papers on the quark model:\\
``Electromagnetic Mass Differences in the Quark Model''~\cite{h1}\\
``Spin $2^+$ Meson Decays in the Quark Model''~\cite{h2}\\
``Dynamical Derivation of Baryon Masses in the Quark Model''~\cite{h3}\\
``Electromagnetic Properties of Hadrons in the Quark Model''~\cite{h4}\\
``Nucleon-Antinucleon Annihilation in the Quark Model''~\cite{h5}\\
``A Dynamical Quark Model for Hadrons.''~\cite{h6}\\
This impressive output makes clear that Hector completely invested in the idea that quarks
are real, not just a mechanism to get group theory results. He did realize that
quarks could be used to get $SU(3)$ and $SU(6)$ results; nonetheless he
was outspoken and unambiguous that quarks are real. This reflected both his
physics judgement and his forthright and enthusiastic personality.

\section{Quantum statistics}\label{quantum statistics}

I will describe three issues related to quantum statistics:\\
(1) Difficulties to make a model with small violations of statistics,\\
(2) Theoretical possibilities for statistics in three space dimensions,\\
(3) Types of experimental tests.\\

\subsection{Difficulty to violate statistics by a small amount}\label{small violations}

The Hamiltonian (and all observables) must commute with permutations of
the identical particles; otherwise they would not be treated in an equivalent way.
This is a very primitive condition. It does not depend on locality, or covariance,
or quantum field theory. In particular, to violate statistics you cannot just add a
small term that violates statistics, say $H=H_0+H_V,$ where $H_V$ is small and
does not commute with permutations, in analogy with the way you could add a
small term that violates CP or other symmetries. A small statistics-violating term
would destroy the indistinguishability of the identical particles. Also, you cannot
introduce a new degree of freedom, like red electrons or blue electrons, because that
would double the pair production cross section. Violating statistics by a small amount
requires something subtle.

\subsection{What quantum mechanics allows}\label{what quantum mechanics allows}

Messiah's ``symmetrization postulate'' (SP),~\cite{mes} that states of several identical particles
are either symmetric or antisymmetric under permutations, is equivalent to stating that
identical particles only occur in one-dimensional representations of the symmetric group.
However, quantum mechanics allows states not obeying the symmetrization postulate;
the symmetrization postulate is an additional assumption that is not a basic principle of
quantum mechanics. To formulate a version of quantum mechanics without the symmetrization
postulate, the concept of ``ray'' must be generalized to the concept of ``generalized ray,'' in
which the phase ambiguity of a ray is replaced by the ambiguity of a unitary representation of an irreducible
representation of the symmetric group. Indeed, the usual phase ambiguity is a unitary representation
of a one-dimensional irreducible representation of the symmetric group. States in inequivalent
representations of the symmetric group cannot be coherent, i.e. cannot interfere, because no
observable connects such states.

This incoherence leads to a primitive superselection rule that Messiah and I
first pointed out~\cite{mes2} and that R. Amado and H. Primakoff~\cite{ama} also emphasized: States in inequivalent
representations of the symmetric group are separated by this superselection rule. Corollaries of this are
that no transitions can occur between SP-obeying and SP-violating states, and no transitions can occur
between bosonic and fermionic states. I emphasize that this superselection rule follows from the
meaning of identical particles, and does not require Lorentz covariance, locality, or other properties
of quantum field theory

In theories with noncommutative spacetime coordinates the Drinfel'd
``twists''~\cite{dri} can convert fermi$\rightarrow$fermi or bose$\rightarrow$bose
transitions to the opposite type, but
cannot produce bose$\rightarrow$fermi transitions or vice versa.

\subsection{Theoretical possibilities for quantum statistics}\label{possible quantum statistics}

Three types of statistics~\cite{dop} can occur in 3 or more space dimensions: parabose statistics of integer order $p$,~\cite{hsg}
parafermi statistics of integer order $p$,~\cite{hsg} and infinite statistics.~\cite{inf} (In fewer than 3 dimensions, fractional statistics
(anyons) can occur.) The rule we all know and love, that bose statistics occur for particles having integer
spin and fermi statistics occur for particles having odd-half integer statistics, also holds in the obvious way
far parabose and parafermi statistics. There is an exception, not yet discovered in nature, the dyon. The dyon
is a composite of a a magnetic monopole, with magnetic charge $g$ and a charged particle with electric
charge $e$. This system has angular momentum stored in its electromagnetic field. If $eg/2 \pi$ is odd,
the dyon violates the spin-statistics connection for the charged particle in 3 dimensions.

\subsubsection{Parastatistics}\label{parastatistics}

Parastatistics theories can be formulated as local relativistic quantum field theories. Parabose (parafermi) theories
of order $p$ can be represented as sums of $p$ bose (fermi) theories with anomalous relative commutation
relations via the Green ansatz. Messiah and I showed that these theories can also be treated without
using the Green ansatz, and that the Green ansatz can be derived for any representation of the Green
trilinear commutation rules that has a Fock-like vacuum state. In each case, $p=1$ corresponds to the
usual bose and fermi theories. These parastatistics theories have gross violations of the usual statistics
and do not provide a model for small violations of statistics.

\subsubsection{Infinite statistics}\label{infinite statistics}

One can find the commutation relations for infinite statistics by averaging the commutation relations for
bose and fermi statistics,
\beee
(1/2)\{[a_k,a_j\dggg]_- + [a_k,a_j\dggg]_+\}=a_ka_l\dggg=\delta_{kl}
\eeee
This algebra is called the Cuntz algebra in the mathematical literature. With the Fock-like vacuum
condition, $a_k|0\rangle =0$, one can calculate all matrix elements of products of annihilation and
creation operators. One does not need a relation with only creation or only annihilation operators.
Further, no such relation can be imposed consistently. The norm of every monomial in the annihilation
operators acting on the vacuum is 1. The Gibbs correction factor is absent here. This statistics is
``quantum Boltzmann'' statistics. All representations of the symmetric group, $S_n$, occur with equal
weights.

To generalize infinite statistics as described by the Cuntz algebra, take a convex sum of the bose
and fermi commutation relations,
\beee
\frac{1+q}{2}[a_k,a\dggg_l]_- +\frac{1-q}{2}[a_k,a\dggg_l]_+
=a_k a\dggg_l-q a\dggg_l a_k=\delta_{kl}.
\eeee
This is called the quon algebra. Again, the quon commutation relations and the Fock vacuum condition
suffice to calculate all matrix elements, and
no relation on $aa$ or $a\dggg a\dggg$ is needed and
none can be imposed, except for $q= \pm 1$. An analog of Wick's theorem holds. In contrast to the bose and
fermi cases, the terms have coefficients that are powers of $q$ rather than the $\pm 1$ that occur in the
bose and fermi cases.~\cite{owgmis}

For a state with $n$ quons, all representation of $S_n$ occur, with weights dependent on $q$. For
$q \rightarrow -1$ the more antisymmetric representations dominate, and for
$q \rightarrow 1$ the more symmetric representations dominate.
At $q =0$ all representations occur with equal weight. Quons give a small violation of statistics
by producing a mixed density matrix. The smallness of the violation comes from the smallness of
the mixture of ``abnormal'' states.

\subsection{Experimental tests of quantum statistics}\label{experimental tests}

There are four main types of tests of quantum statistics:\\
(i) Transitions between anomalous states,\\
(ii) Accumulations in anomalous states,\\
(iii) Deviations from the usual statistical properties of identical particles,\\
(iv) Stability of matter.

The pioneering experiment to search for transitions between anomalous states was carried
out by M. Goldhaber and G. Scharff-Goldhaber~\cite{mgold} in 1947. They pointed out that if the electrons
in beta decay were not quantum-mechanically identical to the electrons in atoms, then
beta decay electrons could fall into the K shell of an atom in violation of the Pauli
exclusion principle. They let the beta decay electrons from a naturally occurring radioactive
source fall on a block of lead and looked for the x-rays from transitions into the K shell.
They did not find such x-rays and found a qualitative bound on violations of the exclusion
principle. E. Ramberg and G. Snow~\cite{ram} repeated this type of experiment running a 30 A current in a
thin copper strip for one month and looking for the slightly displaced x-rays. They had
Avogadro's number on their side and found a null result with high precision. With the
two-body density matrix in the form
\beee
\rho = (1-v_F) \rho_A + v_F\rho_S
\eeee
Ramberg and Snow found the bound $v_F \leq 1.7 \times 10^{-26}$.
This bound was improved by the VIP experiment of Bartalucci, et al~\cite{bar} in (2010), which
gave the bound $v_F \leq 6 \times 10^{-29}$ This improvement came from three factors,
(i) using CCD detectors that give greater sensitivity, (ii) running the experiment underground
with less background, and (iii) taking a longer run which give greater statistics.

High-precision tests of bose-einstein statistics for photons are more difficult because
we don't have bound states with many photons. In 2010, D. English, et al,~\cite{eng} looked for
bose-einstein forbidden two-photon transitions in atomic barium. These transtions are
forbidden by the Landau-Yang bose-einstein selection rule. They found the bound
$\nu$ forbidden/$\nu$ allowed $\leq 4.0 \times 10^{-11}$.

In two papers in 2010, A.P. Balachandran, et al,~\cite{bal} studied the possibility of violations
of statistics in theories with noncommutative space coordinates. They found that the
Drinfel'd ``twists'' can induce bose $\rightarrow$ bose transitions between two-electron states
that obey the exclusion principle. The mass scale for these transitions is well above the
Planck scale, so such transitions cannot be observed at present.

We conclude that high-precision tests have ruled out violations of statistics with techniques
now available. No violations of Lorentz covariance have been found. Superpartners have not
been found. Extra dimensions have not been found. We have high hopes for such discoveries at the LHC.

H\'ector had the wide ranging interests, and the perspicacity, to turn to astrophysics and cosmology
where the great discoveries, dark energy and dark matter, have been made in recent years.

H\'ector had, as did many of us here today, the good luck to live through the wonderful years
in which the standard model of elementary particles was developed. He made significant contributions
to this development. We are all sorry that he is
no longer with us inspire us with his contributions and with his insight in his new interests,
astrophysics and cosmology.

\acknowledgments

I thank Dr.\ Dan-Olof Riska for his hospitality at the
Helsinki Institute for Physics where I prepared this talk.\\

Here is the photo of the Bebek summer institute, as well as a photo of H\'ector taken in 2008.

\begin{figure}[h]
\includegraphics[width=\textwidth]{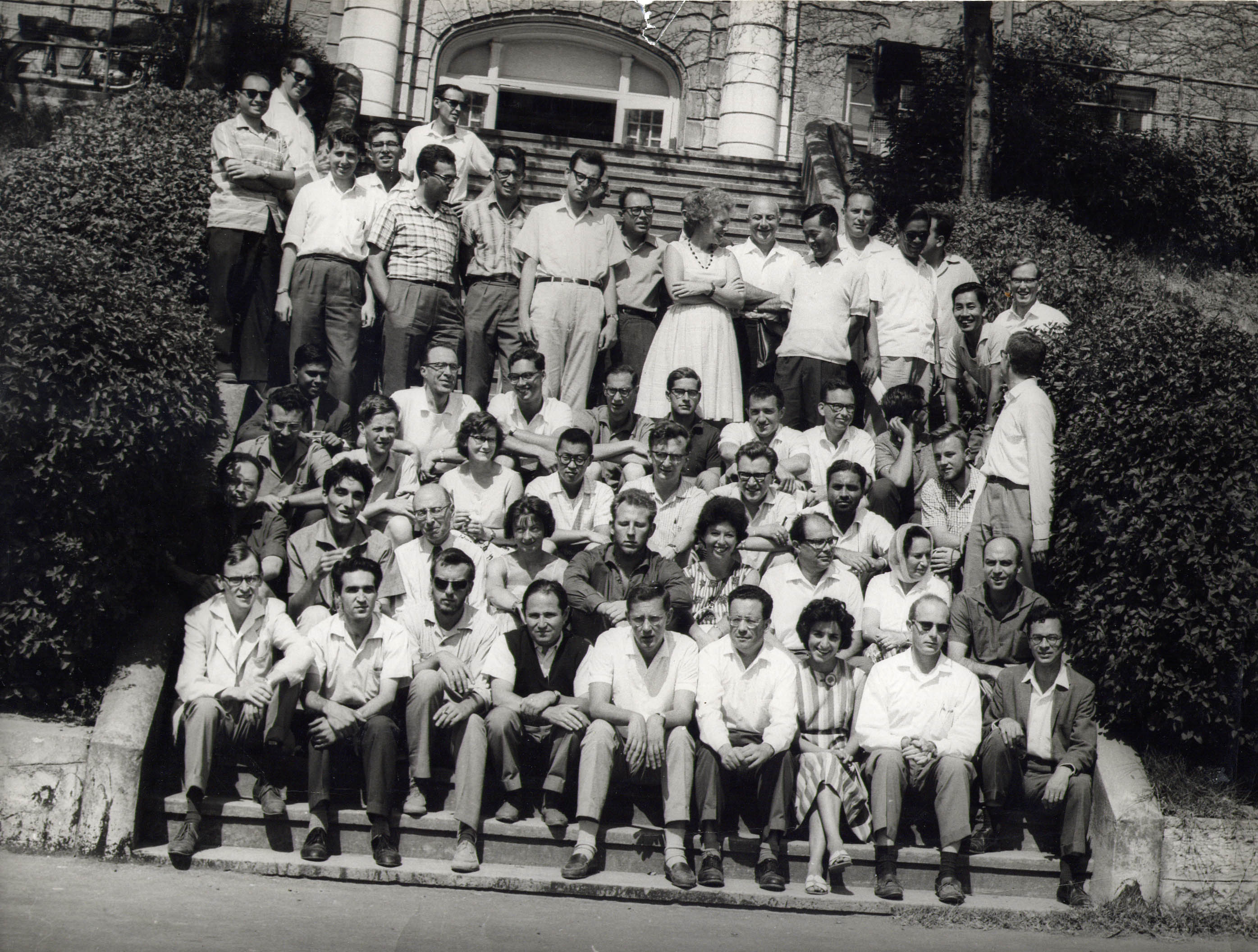}
\includegraphics[width=\textwidth]{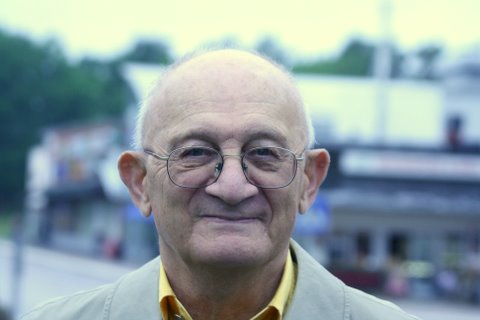}
\end{figure}




\end{document}